\documentclass{aa} 
\usepackage{graphicx}

\begin{document}

\title{Orbital motion in T~Tauri binary systems
\thanks{Based on observations collected at the German-Spanish Astronomical
 Center on Calar Alto, Spain, and at the European Southern Observatory,
 La Silla, Chile.}}

\author{J.\,Woitas\inst{1, 2}
 \and R.\,K\"ohler\inst{3, 2}
 \and Ch.\,Leinert\inst{2}
 }

\offprints{Jens Woitas, \email{woitas@tls-tautenburg.de}}

\institute{Th\"uringer Landessternwarte Tautenburg, Sternwarte 5,
 07778 Tautenburg, Germany 
 \and Max-Planck-Institut f\"ur Astronomie, K\"onigstuhl 17,
 69117 Heidelberg, Germany
 \and Center for Astrophysics and Space Sciences, University of California,
  San Diego, 9500 Gilman Drive, La Jolla, \newline CA 92093-0424, USA}

\date{Received / Accepted}

\abstract{
Using speckle-interferometry we have carried out repeated measurements 
of relative positions for the components of 34 T~Tauri binary systems.
The projected separation of these components is low enough that orbital
motion is expected to be observable within a few years. In most cases
orbital motion has indeed been detected. The observational
data is discussed in a manner similar to Ghez et al.\,(\cite{Ghez95}).
However, we extend their study to a larger number of objects and a much
longer timespan.
The database presented in this paper is valuable for future visual
orbit determinations. It will yield empirical masses for T~Tauri stars
that now are only poorly known.  The available data is however not sufficient
to do this at the present time. Instead, we use short series of orbital data
and statistical distributions of orbital parameters to derive an average
system mass that is independent of theoretical assumptions about the
physics of PMS stars. For our sample this mass is $2.0\,M_{\sun}$ and thus
in the order of magnitude one expects for the mass sum of two T~Tauri
stars. It is also comparable to mass estimates obtained for the same systems
using theoretical PMS evolutionary models. 
}

\maketitle

\section{Introduction}
\label{intro}
The mass is the most fundamental parameter of a star, because to a large
extent it determines its structure and evolution. Therefore, it is a major
problem for our understanding of pre-main sequence (PMS) evolution that
at this time there are no reliable empirical mass determinations for young
stars with $M < 1\,M_{\sun}$. Mass estimates for T~Tauri stars are usually
based on comparisons of their position in the Hertzsprung-Russell diagram
(HRD) with theoretical PMS evolutionary models, which means that they are
affected by the (unknown) uncertainties within these models. Moreover,
it is not possible to rate the quality of different sets of PMS models by
comparison to observational data.\\
For this reason, empirical mass determinations for young stars are highly
desirable. Binary stars offer a unique possibility to do this, because the
system mass is known as soon as the orbital parameters are determined.
There are many binaries among T~Tauri stars in nearby star-forming
regions (SFRs). Most of them have been detected during the last
decade by high-angular resolution surveys in the near infrared (NIR)
(For an overview of this topic, see the review by Mathieu
et al.\,(\cite{Mathieu00}) and references therein.).\\
The first reliable empirical masses of PMS stars were given by Casey et
al.\,(\cite{Casey98}) for the components of the eclipsing double-lined
spectroscopic binary (ESB2) \object{TY\,CrA}. These masses are
$M_1 = 3.16 \pm 0.02\,M_{\odot}$ and $M_2 = 1.64 \pm 0.01\,M_{\odot}$.
The secondary mass is consistent with the predictions of PMS models from
D'Antona \& Mazzitelli\,(\cite{dm94}) and also  Swenson et
al.\,(\cite{Swenson94}). The primary is already close to the main sequence.
The lowest-mass PMS stars with empirically determined masses thus far known
are the components of \object{RX\,J0529.4+0041}. For this ESB2,
Covino et al.\,(\cite{Covino2000}) determined the masses
$M_1 = 1.25 \pm 0.05\,M_{\odot}$ and $M_2 = 0.91 \pm 0.05\,M_{\odot}$.
They concluded that these masses are in good agreement with with the
Baraffe et al.\,(\cite{Baraffe98}) and Swenson et al.\,(\cite{Swenson94})
models, but less consistent with sets of PMS tracks from 
D'Antona \& Mazzitelli\,(\cite{dm94}) and
Palla \& Stahler\,(\cite{Palla99}). Because of the relatively high masses,
these results cannot be used to check the PMS models for
K- or M-dwarfs and objects with masses below the hydrogen burning mass
limit at $\approx 0.075\,M_{\odot}$.\\
In this paper we will follow the approach of Ghez et
al.\,(\cite{Ghez95}, herafter G95). Using NIR speckle interferometry, they
obtained repeated measurements for the relative astrometry of the
components in 20 T~Tauri binary systems.
In this way they showed that in most of these systems, orbital motion
can be determined. From short pieces of orbital data and the statistical
distribution of orbital parameters, they have derived the {\it average}
system mass of $1.7\,M_{\odot}$, which is in the order of magnitude expected
for the mass of two T~Tauri stars.\\
We present similar data for 34 T~Tauri binary systems and in this way
increase the object list and also the observational time base. An overview
of our observations and data reduction is given in Sect.\,\ref{obs}.
The results are presented in Sect.\,\ref{results}, discussed in
Sect.\,\ref{discussion} and summarized in Sect.\,\ref{summary}.

\begin{figure}
\resizebox{\hsize}{!}
{\includegraphics{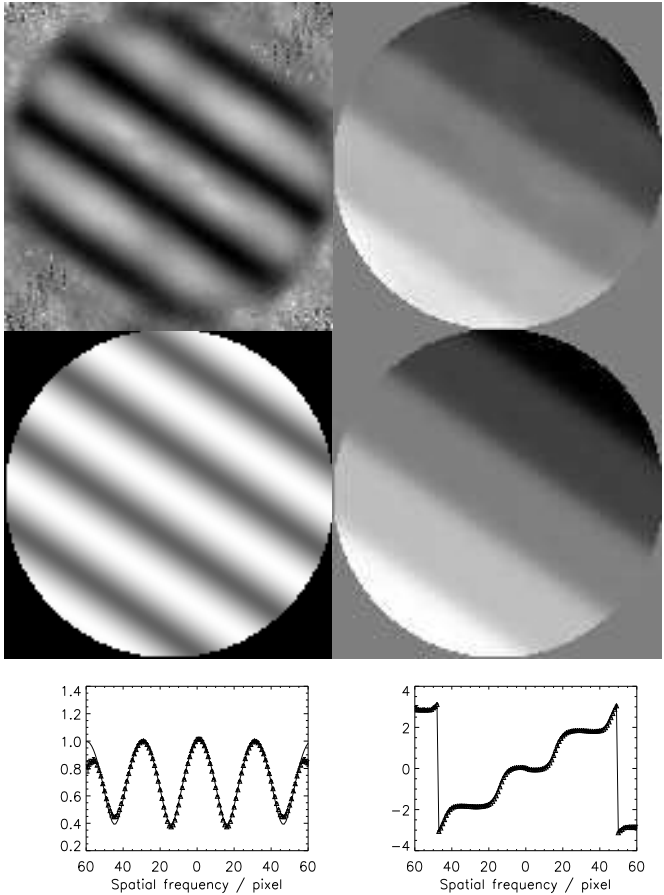}}
\caption{\label{vispha}
 The first row shows modulus (left) and bispectrum-phase (right) of
 the complex visibility for the binary XZ Tau, derived from data obtained on
 29 Sep 1996 at the 3.5\,m-telescope on Calar Alto with the NIR array camera
 MAGIC at $\lambda = 2.2\,\mu\mathrm{m}$. The second row is the modulus
 and phase of an artificial complex visibility that is fitted to the data.
 This fit is indicated in the third row in a one-dimensional projection
 towards the connection line of the components (perpendicular to the
 strip patterns in Fourier space). The circle around the strip patterns
 corresponds to the Nyquist spatial frequency, $7.0\,\mathrm{arcsec}^{-1}$
 for the adopted pixel scale.}
\end{figure}

\section{Observations and data reduction}
\label{obs}

\subsection{The sample}
\label{sample}
We have repeatedly observed 21 systems in Taurus-Auriga detected as
binaries by Leinert et al.\,(\cite{Leinert93}) and K\"ohler \& Leinert
(\cite{Koehler98}) and 11 systems in Scorpius-Centaurus detected by
K\"ohler et al.\,(\cite{Koehler2000}). Furthermore, we re-observed two
binaries found by Ghez et al.\,(\cite{Ghez97a}). These are \object{HM Anon} in
the Chamaeleon\,I association, and \object{HN Lup} in the Lupus SFR.
We combine our data with results taken from literature
(see Sect.\,\ref{database}). In particular twelve systems discussed by G95
are also objects of our study.

\subsection{Observations}
\label{observations}
The objects in Taurus-Auriga were observed with the 3.5\,m-telescope on
Calar Alto. After September 1993 these observations used the NIR
array camera MAGIC. Previous measurements were obtained with a
device for one-dimensional speckle interferometry described by
Leinert \& Haas (\cite{Leinert89}). The observations of
young binaries in southern SFRs were carried out at the ESO New Technology
Telescope (NTT) at La Silla that is also a 3.5\,m-telescope. The instrument
used for these observations was the NIR array camera SHARP I of the
Max-Planck Institute for Extraterrestrial Physics (Hofmann et
al.\,\cite{Hofmann92}). Both cameras are capable of obtaining fast sequences
of short exposures with integration times $\tau\approx 0.1\,\mathrm{s}$,
which is crucial for the applied data reduction process (see
Sect.\,\ref{speckle}).\\ 
Most of the data were obtained in the K-band at
$\lambda~=~2.2\,\mu\mathrm{m}$. Some observations used the J-band at
$\lambda~=~1.25\,\mu\mathrm{m}$ and
the H-band at $\lambda~=~1.65\,\mu\mathrm{m}$. In these cases the main
goal of the observations was to obtain resolved photometry of the components
at those wavelenghts. In the course of data reduction we could however
show that highly precise relative astrometry can also be derived from
observations in J and H (see Sect.\,\ref{params} for the determination
of binary parameters).

\subsection{Speckle interferometry}
\label{speckle}
At the presumed distance of the objects (between 140 and 190\,pc, 
see Sect.\,\ref{relative} and Table\,\ref{distances}) orbital motion
measurable within a few years can only be expected for the closest pairs
with projected separations of $\mathrm{d}\le 0\farcs5$. Therefore, high
angular resolution techniques are necessary that overcome the effects of
atmospheric turbulence and yield the diffraction-limited information about
the objects. For this purpose, we have used speckle interferometry.\\
Sequences of $\approx 1000$ short exposures ($\tau \approx 0.1\,\mathrm{s}$)
are taken for the object and a nearby point source, the reference
star. The integration time is shorter than the coherence time of the
turbulent atmosphere, so the turbulence is ``frozen'', and the images
are noisy, but principally diffraction limited. After Fourier transforming
these ``data cubes'', the power spectrum of the image is deconvolved
with that of the reference star to obtain the modulus of the complex
visibility. The phase is reconstructed using the Knox-Thompson algorithm
(Knox \& Thompson\,\cite{Knox74}) and the bispectrum method
(Lohmann et al.\,\cite{Lohmann83}). The complex visibility is the
Fourier transform of the object brightness distribution. For a sufficiently
bright object it will contain the diffraction-limited information.

\subsection{Determination of binary parameters}
\label{params}
Modulus and phase of the complex visibility are characteristic strip patterns
for a binary, as shown in Fig.\,\ref{vispha} (first row) for the
\object{XZ Tau} system. The position angle of the companion is orientated
perpendicular to these patterns and towards higher values of phase.
The distance between two strips is inversely proportional to the projected
separation of the components, and the amplitude of the patterns denotes
the flux ratio. Position angle, projected separation and flux ratio are
determined by constructing an artificial complex visibility
from a set of these parameters (second row in Fig.\,\ref{vispha}) and
fitting it to the data (third row in Fig.\,\ref{vispha}).
This fit uses the {\tt Amoeba} algorithm (Press et al.\,\cite{Press94}).
The errors of the binary parameters are estimated by applying this fitting
procedure to different subsets of the data.

\subsection{Pixel scale and detector orientation}
\label{skala}
To obtain position angle and projected separation as astronomical
quantities they must be transformed from the array onto the sky
by calibrating them with pixel scale and orientation of the detector.
Minimizing the uncertainties that result from this calibration is crucial
for a highly accurate determination of the components' relative
astrometry that is proposed here. We determine pixel scale and detector
orientation from astrometric fits to images of the Orion Trapezium
cluster where precise astrometry has been derived by McCaughrean \&
Stauffer (\cite{McCaughrean94}). Typical errors are $0.1^{\circ}$
for the detector orientation and $10^{-4}\,\mathrm{arcsec/pixel}$
for the pixel scale that is $\approx 0\farcs07/\mathrm{pixel}$ for MAGIC
and $\approx 0\farcs05/\mathrm{pixel}$ for SHARP I.\\
This precise calibration exists for all observations obtained by the authors
since July 1995. For calibrating data from previous observing runs we
used visual binary stars with well-known orbits. We observed these stars
afterwards and calibrated their separation and position angle with the help
of the Trapezium cluster. Thus, we have placed all our (two-dimensional)
speckle observations into a consistent system of pixel scale and detector
orientation.

\subsection{The database}
\label{database}
In Table\,\ref{obs-table} the derived position angles, projected
separations and flux ratios are presented. For the subsequent analysis
(Sect.\,\ref{results}) the relative positions are combined with data taken
from literature. Twelve of our systems
have also been discussed by G95. For some binaries there are additional
measurements obtained by HST imaging, the HST Fine Guidance Sensors or
adaptive optics (see references in Table\,\ref{obs-table}). 
The finding that our relative astrometry fits very well with that obtained
by other groups further supports the idea that the position angles and
projected separations are as precise as is indicated by the errors given
in  Table\,\ref{obs-table}.  

\begin{table}
\caption{\label{distances} Adopted distances to stars in nearby SFRs}
\begin{tabular}{lll}
 & & \\ \hline
SFR & Distance\,[pc] & Reference \\ \hline
Taurus-Auriga & 142 $\pm$ 14 & Wichmann et al.\,(\cite{Wichmann98}) \\
Scorpius-Centaurus & 145 $\pm$ 2 & de Zeeuw et al.\,(\cite{deZeeuw99}) \\
Chamaeleon\,I & 160 $\pm$ 17 & Wichmann et al.\,(\cite{Wichmann98}) \\
Lupus & 190 $\pm$ 27 &  Wichmann et al.\,(\cite{Wichmann98}) \\
\hline
\end{tabular}
\end{table}

\section{Results}
\label{results}

\subsection{Relative motion of the components}
\label{relative}
\begin{figure*}
 \resizebox{\hsize}{!}{\includegraphics{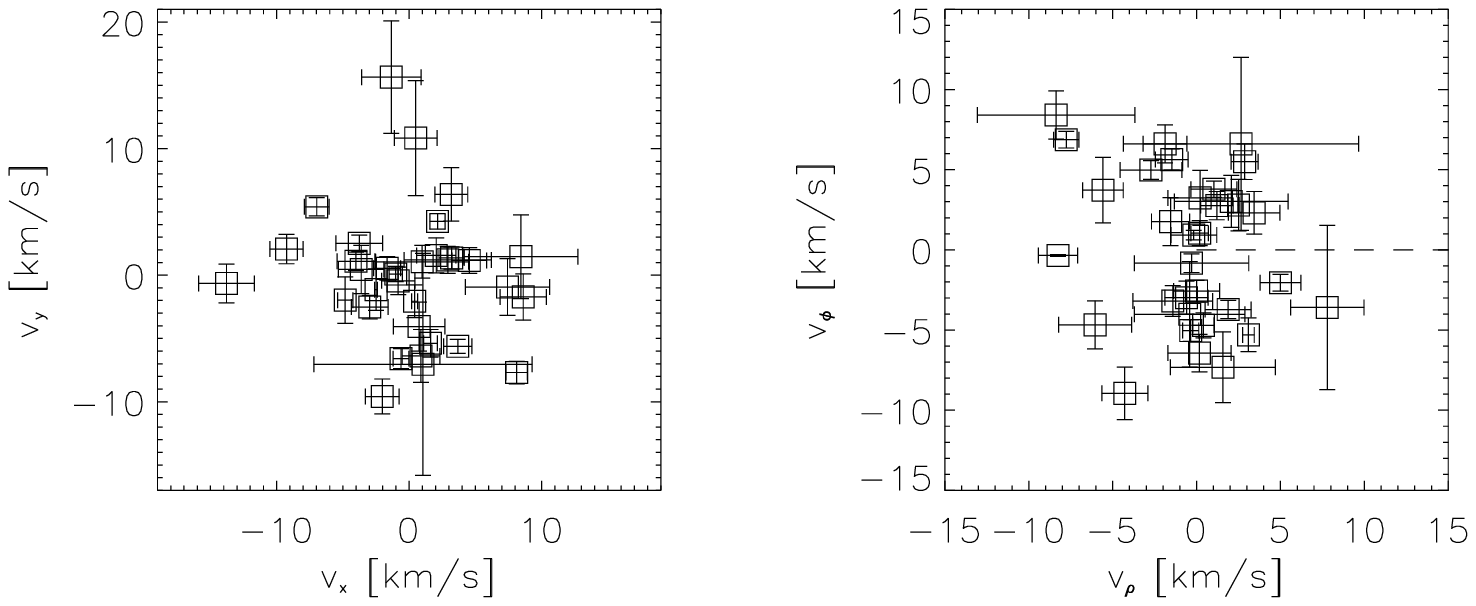}}
 \caption{\label{vrel} Relative velocities of the components in T~Tauri
  binary systems in cartesian coordinates (left panel) and polar
  coordinates (right panel). The dashed line in
  the right panel points towards the locus of Herbig-Haro objects that
  is far outside this plot. RX\,J1546.1-2804 is not plotted in this
  figure. Its locus is out of the picture in both panels. This plot
  is similar to Fig.\,3 in G95.}
\end{figure*}

In Fig.\,\ref{orbitplots} the relative positions of the components at 
different epochs are shown in cartesian and polar coordinates. These plots
are only given for the 23 out of our 34 systems for which there are at least
three data points. To obtain a simple approximation of the relative velocity
we applied weighted linear fits to this data. For the 11 systems with
only two observations we simply connect the two data points and derive the
error of the relative velocity from the uncertainties of the two 
separations and position angles.\\
For a quantitative analysis, the relative velocities must be transformed
to an absolute length scale. This requires knowing the distances of the
discussed objects. We adopt distances to the SFRs that are the mean of all
Hipparcos distances derived for members of the respective association.
The values and references are given in Table\,\ref{distances}.
There remains, however, an uncertainty, because distances of individual
objects may be different from these mean values. To take this into account, we
assume that the radial diameters of the SFRs are as large as their
projected diameters on the sky. The latter quantity can be estimated
to be $\approx 20^{\circ}$ for Taurus-Auriga (see Fig.\,1 in K\"ohler
\& Leinert \cite{Koehler98}) as well as Scorpius-Centaurus (see Fig.\,1
in K\"ohler et al.\,\cite{Koehler2000}). Concerning the mean distances
from Table\,\ref{distances}, this corresponds to a diameter of 50\,pc.
Thus we will assume $\pm 25\,\mathrm{pc}$ as the uncertainty for the
distance of an individual system, which is an upper limit: More than two
thirds of the stars will be within $\pm$ 15\,pc for an even distribution.\\
The velocities of the companions relative to the primaries derived by
applying the assumed distances are given in Table\,\ref{vrel-table}.
They are also plotted in Fig.\,\ref{vrel} in cartesian and polar
coordinates (similar to Fig.\,3 in G95). Our measurements can only cover the
projection of motion onto the sky, so the $v_{\rho}$ are given with respect to
the main component, not to the observer. The adopted $v$ is
the mean of the respective values derived from the fits in cartesian
and polar coordinates. In 3 out of 34 systems $v$ is different from zero
on the 3$\sigma$ level, in 9 systems on the 2$\sigma$ level and in
18 systems on the $1\sigma$ level. Thus, we are fairly confident that
there really is relative motion of the components in most systems.
 
\begin{table*}
\caption{\label{vrel-table} Projected relative velocities of the companions
 with respect to the primaries in cartesian and polar coordinates. The
 adopted $v$ is the mean of the total velocities derived from $(v_x, v_y)$
 and $(v_{\rho}, v_{\phi})$. Note that $v_{\rho}$ is given with respect
 to the main component and not relative to the observer.}
\begin{tabular}{lllllll}
\hline
System & $v_x$[km/s] & $v_y$[km/s] & $v_{\rho}$[km/s] & $v_{\phi}$[km/s] &
  v[km/s] & $\bar{d}$[AU] \\ \hline
V 773 Tau & -1.35 $\pm$ 2.47 & 15.64 $\pm$ 6.87 & -8.37 $\pm$ 6.82 &
    8.41 $\pm$ 2.83 &  13.78 $\pm$ 7.34 & 14.9 \\
LkCa 3    & 4.54 $\pm$ 1.92 & 1.16 $\pm$ 1.07 & -0.60 $\pm$ 1.49 & 
   -2.98 $\pm$ 1.62 &  3.86 $\pm$ 2.20 &  68.7 \\
FO Tau    & 3.67 $\pm$ 1.63 & -5.62 $\pm$ 1.52 & -1.48 $\pm$ 1.36 &
    5.63 $\pm$ 1.72 &  6.27 $\pm$ 2.21 &  22.3 \\
CZ Tau    & -4.83 $\pm$ 1.46 & -1.97 $\pm$ 2.22 & -2.72 $\pm$ 2.27 &
    4.98 $\pm$ 1.43 &  5.45 $\pm$ 2.67 & 45.3 \\
FS Tau    & -9.26 $\pm$ 2.87 & 2.06 $\pm$ 1.44 & -1.88 $\pm$ 1.54 &
    6.61 $\pm$ 2.77 &  8.18 $\pm$ 3.19 & 35.6 \\
FW Tau    & 8.11 $\pm$ 2.12 & -7.69 $\pm$ 2.22 & -7.77 $\pm$ 2.24 &
    6.87 $\pm$ 1.67 & 10.77 $\pm$ 2.93 &  15.8 \\
LkH$\alpha$ 331 & -3.77 $\pm$ 2.21 &  2.51 $\pm$ 1.04 & -1.55 $\pm$ 1.65 &
    1.76 $\pm$ 1.77 &  3.44 $\pm$ 2.43 & 40.3 \\
XZ Tau & 2.14 $\pm$ 1.03 & 4.25 $\pm$ 1.32 &  -0.35 $\pm$ 0.50 &
   -5.03 $\pm$ 1.52 &  4.90 $\pm$ 1.64 &  43.2 \\
HK Tau G2 & 0.41 $\pm$ 0.34 & -2.06 $\pm$ 1.92 &  5.00 $\pm$ 1.80 &
   -2.05 $\pm$ 1.92 &  3.75 $\pm$ 1.95 &  26.3 \\
GG Tau Aa & -0.62 $\pm$ 0.69 & -6.59 $\pm$ 1.82 & 0.41 $\pm$ 0.72 &
   -4.71 $\pm$ 1.80 &  5.67 $\pm$ 1.94 &  35.8 \\
UZ Tau/w & 3.20 $\pm$ 1.31 & 1.36 $\pm$ 1.24 &  1.21 $\pm$ 1.32 &
    2.76 $\pm$ 1.24 &  3.24 $\pm$ 1.81 &  50.9 \\
GH Tau & 8.61 $\pm$ 2.96 & -2.97 $\pm$ 2.14 & -1.42 $\pm$ 3.18 &
   -3.19 $\pm$ 1.92 &  6.30 $\pm$ 3.68 &  45.0 \\
Elias 12 & -13.79 $\pm$ 3.95 & -0.65 $\pm$ 1.71 & -6.05 $\pm$ 3.67 &
   -4.68 $\pm$ 2.56 & 10.73 $\pm$ 4.39 & 49.3 \\
IS Tau & -4.55 $\pm$ 2.35 & 0.45 $\pm$ 1.92 & 0.21 $\pm$ 1.93 &
   3.24 $\pm$ 2.37  & 3.91 $\pm$ 3.04  & 31.7 \\
IW Tau & -2.50 $\pm$ 0.55 & -1.14 $\pm$ 1.91 & 2.50 $\pm$ 0.47 &
   2.79 $\pm$ 1.80  & 3.24 $\pm$ 1.97 & 39.6 \\
LkH$\alpha$ 332/G2 & 2.31 $\pm$ 1.80 & 7.68 $\pm$ 3.23 &  -5.58 $\pm$ 2.46 &
   3.72 $\pm$ 2.07  & 7.36 $\pm$ 3.45 & 36.5 \\
LkH$\alpha$ 332/G1 & -1.66 $\pm$ 1.62 & 0.54 $\pm$ 1.23 &  2.86 $\pm$ 1.06 &
   5.50 $\pm$ 1.76  & 3.97 $\pm$ 2.04 & 32.2 \\
LkH$\alpha$ 332 & 2.91 $\pm$ 3.54 & 1.21 $\pm$ 1.20 &  -0.30 $\pm$ 3.69 & 
  -0.83 $\pm$ 0.60  & 2.02 $\pm$ 3.74 & 47.0 \\
BD+26\,718B Aa & -6.98 $\pm$ 2.14 & 5.40 $\pm$ 1.66 &  -8.26 $\pm$ 2.70 & 
  -0.35 $\pm$ 0.12  & 8.55 $\pm$ 2.71 & 67.5 \\
BD+26\,718B Bb & 0.98 $\pm$ 3.60 & 1.06 $\pm$ 1.51 &  2.07 $\pm$ 3.48 &
   3.02 $\pm$ 1.84  & 2.55 $\pm$ 3.92 & 23.3 \\
BD+17\,724B & 0.49 $\pm$ 1.64 & 10.83 $\pm$ 6.20 & -0.41 $\pm$ 3.88 &
  -4.02 $\pm$ 5.15  & 7.44 $\pm$ 6.43 & 12.8 \\
NTTS\,155808-2219 & 0.75 $\pm$ 2.64 & -4.06 $\pm$ 2.06 & 3.43 $\pm$ 2.13 &
  2.31 $\pm$ 1.72 & 4.13 $\pm$ 3.05 & 29.4 \\
NTTS\,155913-2233 & 1.60 $\pm$ 0.80 & -5.38 $\pm$ 2.03 & 3.09 $\pm$ 0.90 &
   -5.30 $\pm$ 1.85 &  5.87 $\pm$ 2.11 &  43.3 \\
NTTS\,160735-1857 & 0.89 $\pm$ 2.51 & -6.38 $\pm$ 2.24 & 0.17 $\pm$ 1.91 &
   -6.44 $\pm$ 2.28 & 6.44 $\pm$ 3.18 & 43.4 \\
NTTS\,160946-1851 & -1.02 $\pm$ 0.64 & 0.09 $\pm$ 0.61 & 0.18 $\pm$ 0.65 &
    1.04 $\pm$ 0.64 & 1.04 $\pm$ 0.90 & 30.5 \\
HM Anon     & 1.04 $\pm$ 9.33 & -7.04 $\pm$ 8.94 & -2.65 $\pm$ 7.43 &
    6.61 $\pm$ 6.42 & 7.12 $\pm$ 11.37 & 42.1 \\
HN Lup      & 7.42 $\pm$ 3.30 & -0.94 $\pm$ 3.21 & 1.57 $\pm$ 3.33 &
    -7.32 $\pm$ 3.15 & 7.48 $\pm$ 4.60 & 46.3 \\
RX\,J1546.1-2804 & 22.34 $\pm$ 1.58 & 2.86 $\pm$ 4.86 & -3.77 $\pm$ 1.84 &
    -24.40 $\pm$ 7.21 & 23.60 $\pm$ 6.28 & 14.3 \\
RX\,J1549.3-2600 & -0.83 $\pm$ 1.33 & -0.45 $\pm$ 1.23 & -0.17 $\pm$ 1.40 &
     0.92 $\pm$ 0.46 & 0.94 $\pm$ 1.64 & 23.7 \\
RX\,J1600.5-2027 & -4.02 $\pm$ 1.61 & 1.10 $\pm$ 1.42 & 1.88 $\pm$ 1.69 &
     -3.72 $\pm$ 1.21 & 4.17 $\pm$ 2.12 & 28.2 \\
RX\,J1601.7-2049 & 2.04 $\pm$ 1.75 & 1.56 $\pm$ 1.73 & 0.00 $\pm$ 1.37 &
    -2.57 $\pm$ 1.77 & 2.57 $\pm$ 2.11 & 29.7 \\
RX\,J1601.8-2445 & 8.43 $\pm$ 4.55 & 1.46 $\pm$ 4.76 & 7.79 $\pm$ 3.53 &
  -3.59 $\pm$ 5.74 & 8.56 $\pm$ 6.66 & 13.3 \\
RX\,J1603.9-2031B & -2.03 $\pm$ 2.93 & -9.58 $\pm$ 1.73 & -4.28 $\pm$ 2.11 &
  -8.95 $\pm$ 3.19 & 9.86 $\pm$ 3.61 & 15.8 \\
RX\,J1604.3-2130B & -2.98 $\pm$ 1.82 & -2.52 $\pm$ 1.43 & 1.03 $\pm$ 1.55 &
  3.78 $\pm$ 1.67 & 3.91 $\pm$ 2.13 & 12.3 \\
\hline
\end{tabular}
\end{table*}

\subsection{Origin of the relative motion}
\label{origin}
We now examine the origin of this relative motion. For this purpose,
we must discriminate orbital motion from an apparent relative
motion that can be caused by the proper motion of a T~Tauri star with
respect to a background star projected by chance or by the proper motions
of two T~Tauri stars projected by chance. One has further to consider that
``companions'' to T~Tauri stars detected with only one observation
in one filter are not necessarily stellar and may be Herbig-Haro
objects. We will also examine the possible influence of unresolved
additional components on the observed motion.

\subsubsection{Orbital motion}
\label{orbital}

\begin{table}
\caption{\label{vsep-table} Distribution of the companions with respect to
 certain relative velocities and separations}
\begin{tabular}{lll}
 & & \\ \hline
 & $\mathrm{d} < 25\,\mathrm{AU}$ & $\mathrm{d} \ge 25\,\mathrm{AU}$ \\ \hline
 $v\ge 5\,\mathrm{km}/\mathrm{s}$ & $7\pm 2.6$ & $11\pm 3.3$ \\
 $v< 5\,\mathrm{km}/\mathrm{s}$ & $3\pm 1.7$ & $13\pm 3.6$ \\ 
\hline
\end{tabular}
\end{table}

\begin{figure}
\resizebox{\hsize}{!}{\includegraphics{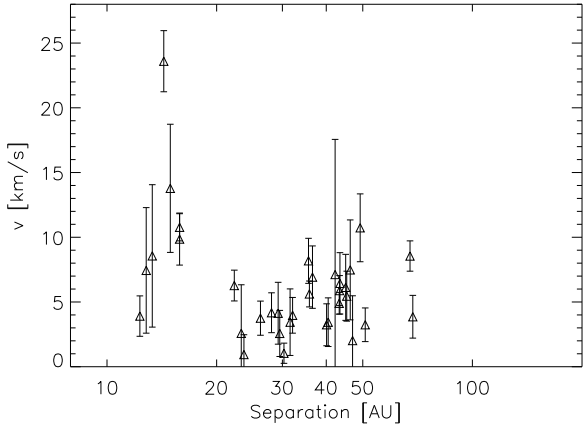}}
\caption{\label{vsep-plot} Relative velocities of the components as
 function of their mean projected separation.}
\end{figure} 

First we will derive an upper limit for relative velocities caused
by orbital motion. That limit is given by the condition that in the case
of orbital motion the kinetic energy of the system is less than its
(negative) potential energy and is equal to it in the extreme case of 
a parabolic orbit, i.\,e.

\begin{equation}
v^2\le\frac{2GM_{\mathrm{Sys}}}{r},
\label{v-limit}
\end{equation}

where $r$ is the instantaneous separation of the components. \\
We adopt $M_{\mathrm{TTS}} \le 2 M_{\sun}$ as upper mass limit for {\it one}
T~Tauri star (Hartmann\,\cite{Hartmann98}) and take the mean of the measured
projected separations as estimate for $r$. There is only one companion
with a relative velocity that is larger than the value derived from
Eq.\,\ref{v-limit}, namely that of \object{RX\,J1546.1-2804}. The
relative velocity of this companion is, however, still consistent with
orbital motion, considering the $1\sigma$ error in its $v$. The lower limit
for $v$ is zero because in short pieces of orbit as discussed here the
orbital motion may occur purely radial to the observer. Thus, the large
majority of the velocities from Table\,\ref{vrel-table} are consistent with
orbital motion.\\
Furthermore, it is interesting to examine the relationship between
separation and relative velocity. In the special case of a circular orbit
observed face-on this relation will be

\begin{equation}
v\propto \frac{1}{\sqrt{a}},
\label{v_prop_a}
\end{equation}

for the orbital velocity $v$ and the semimajor axis $a$. Eq.\,\ref{v_prop_a}
will be modified by projection effects and the actual orbital parameters,
and the unknown system mass is an additional parameter in this relation.
However, close companions should, on average, move faster than distant ones.
In Fig.\,\ref{vsep-plot} the measured projected velocities, $v$, are plotted
as a function of the components' projected separation, $\mathrm{d}$.
The correlation is weak for the reasons mentioned above, but there is at
least a tendency to fulfil the prediction of Eq.\,\ref{v_prop_a}. Among the
closest companions with $d< 25\,\mathrm{AU}$, relative velocities with
$v\ge 5\,\mathrm{km}/\mathrm{s}$ are more frequent than for the companions
with larger separations (Table\,\ref{vsep-table}). \\
We conclude from this section that the observed relative velocities 
in almost all cases are not in contradiction to orbital motion. For a final
classification, other possible origins of the relative motion must be
considered.

\subsubsection{Chance projected background stars}
\label{background}
A background star projected by chance will probably be located at a much
larger distance than the observed T~Tauri star, so its proper motion can
be neglected. The measured relative motion is thus expected to be
the proper motion of the T~Tauri star. In Taurus-Auriga there are
subgroups with different directions of proper motion and a mean proper
motion of $15.8\,\mathrm{km}/\mathrm{s}$ (Jones \& Herbig\,\cite{Jones79}).
For Scorpius-Centaurus,
de Zeeuw et al.\,(\cite{deZeeuw99}) give a mean proper motion of
$17.6\,\mathrm{km}/\mathrm{s}$ with a mean direction of
$v_x = -17.4\,\mathrm{km}/\mathrm{s}$ and
$v_y = -2.8\,\mathrm{km}/\mathrm{s}$. For both SFRs the distances given in
Table\,\ref{distances} were adopted. There are only two systems with
relative velocities in this order of magnitude, namely \object{V 773 Tau}
($v = 13.78\pm 7.34\,\mathrm{km}/\mathrm{s}$) and \object{RX\,J1546.1-2804}
($v = 23.60\pm 6.28\,\mathrm{km}/\mathrm{s}$). \\
For \object{V 773 Tau}, the {\it Hipparcos} catalogue gives proper motions of
$\mu_x = -24.89 \pm 1.89\,\mathrm{mas/yr}$ and 
$\mu_y = 0.65 \pm 2.83\,\mathrm{mas/yr}$. The resultant
$v = 16.68\pm 2.28\,\mathrm{km}/\mathrm{s}$ is comparable with the
observed relative velocity of the components, but the proper motion
of \object{V 773 Tau} happens almost only in declination (X in Fig.
\ref{orbitplots}) which is contradictory to our observations.
Furthermore, for \object{V 773 Tau} the relative motion is still
explainable with orbital motion, considering the upper limit derived
in Sect.\,\ref{orbital}.\\
In the case of \object{RX\,J1546.1-2804} there are no {\it Hipparcos}
data for this individual object, so we adopt the mean proper motion for the
OB association Upper Scorpius given by de Zeeuw et al.\,(\cite{deZeeuw99}).
These values are $\mu = 17.7\,\mathrm{km}/\mathrm{s}$,
$\mu_x = -17.4\,\mathrm{km}/\mathrm{s}$ and
$\mu_y = -2.8\,\mathrm{km}/\mathrm{s}$ with (formal) errors
of $\approx 0.1\,\mathrm{km}/\mathrm{s}$. The respective values
from Table\,\ref{vrel-table} are $v = 23.6\pm 6.3\,\mathrm{km}/\mathrm{s}$,
$v_x = 22.3\pm 1.6\,\mathrm{km}/\mathrm{s}$ and
$v_y = 2.9\pm 4.9\,\mathrm{km}/\mathrm{s}$. This is a close correspondance,
given that the the proper motion presented by de Zeeuw et
al.\,(\cite{deZeeuw99}) is not for this single object, but for the
whole association. Note that $\vec{\mu}$ and $\vec{v}$ must be antiparallel
if the companion is a chance-projected background star, because in
that case $\vec{\mu}$ is the motion of the primary with respect to the
``companion''. Another argument that supports the idea that
the companion of \object{RX\,J1546.1-2804} is a chance-projected
background star is that its relative motion is above the limit
given by Eq.\,\ref{v-limit}. It is, however, still consistent with orbital
motion at the $1\sigma$ level.\\
One must take into account that in both cases the measured
projected separations are $d\approx 0\farcs1$, which makes any chance
projections very unlikely (see Sect.\,\ref{chance} for the probability
of chance projections). Thus, one must consider other origins of these
high relative velocities. One solution may be that the distances
of individual systems are largely different from the adopted values
(Table\,\ref{distances}) for the SFRs as a whole.
The unusually high derived values for relative velocities would then
be caused by overestimating the objects' distances. Another possibile
explanation, namely the presence of unresolved additional companions,
will be discussed in Sect.\,\ref{unresolved}.\\
The companion of \object{RX\,J1546.1-2804} may be such a chance-projected
background star. In the case of \object{V 773 Tau} this seems to be
unlikely because the direction of proper motion does not match, however,
the high relative velocity of the components remains problematic.
Further observations of these systems will be necessary to determine whether
there is a curvature in the relative motion which would undoubtedly classify
it as orbital motion. In general, chance-projected background stars are not
frequent among close visual companions to T~Tauri stars.

\subsubsection{Chance projection of two T~Tauri stars}
\label{chance}
There may also be companions that are in fact objects projected by chance
belonging to the same SFR as the ``primary''. In that case both
components would have roughly the same proper motion. Any
{\it relative} motion would then be caused by the velocity
dispersion within the respective SFR or subgroup, as was
mentioned by G95. These velocity dispersions are between 1 and 2
$\mathrm{km}/\mathrm{s}$ in each coordinate for different subgroups in
Taurus-Auriga (Jones \& Herbig\,\cite{Jones79}).
For Scorpius-Centaurus, de Zeeuw et al.\,(\cite{deZeeuw99}) formally
derived a velocity dispersion of only $0.1\mathrm{km}/\mathrm{s}$.
One cannot distinguish this kind of relative motion from slow orbital motion
by analyzing relative velocities.\\
It is, however, not probable that chance projections within the same SFR
are a frequent phenomenon, because of the low stellar density in the SFRs
discussed here. Leinert et al.\,(\cite{Leinert93}) concluded that there
are less than $4\cdot 10^{-5}$ objects brighter than
$\mathrm{K} = 12\,\mathrm{mag}$ per $\mathrm{arcsec}^2$ in Taurus-Auriga.
This includes association members and field stars.
All companions discussed here are brighter than
$\mathrm{K} = 12\,\mathrm{mag}$ and have projected separations of less than
$0\farcs5$. The mean number of chance-projected companions within a radius
of $0\farcs5$ around our 21 objects in Taurus-Auriga is thus

\begin{equation}
N_{\mathrm{proj}} = 21\cdot 0.5^2\cdot\pi\cdot 4\cdot 10^{-5} =
 6.6\cdot 10^{-4}.
\label{nproj}
\end{equation}

For Scorpius-Centaurus, K\"ohler et al.\,(\cite{Koehler2000}) derived
a stellar density of $(6.64\pm 0.45)\cdot 10^{-4}\,\mathrm{arcsec}^{-2}$
from star counts in the vicinity of X-ray selected T~Tauri stars. 
Applying Eq.\,\ref{nproj} yields a mean number of $5.7\cdot 10^{-3}$
chance-projected companions around 11 objects in Scorpius-Centaurus.
These values are low enough to rule out that chance projections within the
same SFR cause the detection of an appreciable number of close ``companions''.
The low observed relative velocities in some systems are probably due to the
fact that we can only detect the projected motion on the sky.
If most of the orbital motion happens radial to the observer in the time
span covered by our data it will not be measurable by our observations.

\subsubsection{Herbig-Haro objects}
\label{herbig-haro}
G95 raised the question of whether an appreciable number of companions of
young stars detected with only one observation in one broad band filter
may in fact not be stellar, but rather condensations in gaseous nebulae.
Such companions would appear as Herbig-Haro objects that are driven by strong
winds of active T~Tauri stars. They usually have
velocities of some $100\,\mathrm{km}/\mathrm{s}$ (e.\,g.
Schwartz\,\cite{Schwartz83}). Moreover, their relative motion with
respect to their driving source can be only radial, not tangential.
Thus, the locus of Herbig-Haro objects in Fig.\,\ref{vrel} (right panel)
is in the direction of the dashed line in this figure, but far outside
the plot. As was reported by G95, we find no system where the observed
relative velocity is consistent with these restrictions.\\
Furthermore, in the case of \object{HV\,Tau}, where companion C has been
declared as a Herbig-Haro object, we could show that in fact it is a
stellar companion (Woitas \& Leinert\,\cite{Woitas98}).

\subsubsection{Apparent relative motion caused by unresolved companions}
\label{unresolved}
Another problem with observed relative velocities in multiple systems has
been pointed out by G95. If the main component has an additional unresolved
companion, orbital motion in this close pair can shift the photocenter of the
``primary''. This will be misinterpreted as motion of the visual secondary
if only relative astrometry is measured, as is the case here. G95 have
used this effect to explain the surprisingly high relative velocity in the
\object{Elias~12} system where the visual companion is moving with
$10.73\,\mathrm{km}/\mathrm{s}$ at a separation of $49.3\,\mathrm{AU}$
(Table\,\ref{vrel-table}). This is only slightly below the upper limit
for orbital motion, which is $12.0\,\mathrm{km}/\mathrm{s}$ in this case
(Eq.\,\ref{v-limit}). In this system, the primary has another
companion with a separation of $\mathrm{d} = 23\,\mathrm{mas}$, detected by
Simon et al. (\cite{Simon95}) using lunar occultations. If the system mass
of the close pair is assumed to be $1M_{\sun}$ and a relation
$<\mathrm{d}> = 0.95\,a$ between the mean projected separation
$<\mathrm{d}>$ and the semi-major axis $a$
(Leinert et al.\,\cite{Leinert93}) is adopted, one would expect the period
of the close pair to be $P\approx 6.4\,\mathrm{yr}$. This is roughly
the same timespan that is covered by our observations
(Table\,\ref{obs-table}). The shift of the photocenter of the close
pair with respect to the visual secondary will cancel, and thus the derived
secondary's velocity is not affected by the tertiary.\\
Also in the \object{V 773 Tau} system, where we have noticed an unusually
high relative velocity of the visual secondary (Sect.\,\ref{background})
there is an additional
spectroscopic companion (Welty\,\cite{Welty95}). The period of this close
pair is 51.1 days, so possible shifts of the photocenter are less than
2\,mas and thus not measurable by our observations.\\
A candidate for a system where the observed relative motion may be
influenced by an unresolved tertiary is \object{BD+26\,718B} Aa. In this
system $v = 8.55\,\mathrm{km}/\mathrm{s}$ at a separation of
$67.5\,\mathrm{AU}$. Similar to \object{Elias~12}, this value is close to
the upper limit for orbital motion ($v\le 10.2\,\mathrm{km}/\mathrm{s}$
from Eq.\,\ref{v-limit}), but also far below the relative velocity expected
for a background star projected by chance (Sect.\,\ref{background}).
 
\subsubsection{Conclusions}
The observed relative velocties can be explained by orbital motion.
Together with the result that other origins of relative motion can be ruled
out at a high confidence level for most systems, we conclude that the
observed motion is orbital motion in nearly all binaries discussed.\\

\subsection{Estimation of an empirical average system mass}
\label{systemmass}
For all binary systems discussed here the available portions of the orbit
are too short to calculate orbital parameters. The results presented
in Table\,\ref{obs-table}, however, remain valuable for future orbit
determinations that will yield empirical masses for T~Tauri binary systems.
Furthermore, it is already possible to estimate an average system mass
from this database. This average mass is not dependent on theoretical
assumptions about the physics of PMS evolution and should therefore be a
reliable empirical estimation of the masses of T~Tauri stars. To derive this
mass we follow the approach of G95, but improve it in some important
aspects.

First, we write Kepler's third law in the natural units
$M/M_{\sun}$, $a/1\,\mathrm{AU}$ and $P/1\,\mathrm{yr}$:
\begin{equation}
M_{\mathrm{Sys}}  = \frac{a^3}{P^2}
\label{kepler3}
\end{equation}
and assume a face-on circular orbit where the total orbital motion happens
tangential to the observer and therefore equals the observed velocity:
\begin{equation}
v_{\rm face-on, circ} = \frac{2\pi a}{P}.
\label{kepler-circ}
\end{equation}
In this special case ($i=0^{\circ}$, $\mathrm{e}=0$) Eq.\,\ref{kepler3}
becomes
\begin{equation}
M_{\mathrm{Sys}} = \frac{v_{\rm face-on, circ}^2\,a}{4\pi^2}.
\label{msum}
\end{equation}

Unfortunately, we don't know $a$ or $v_{\rm face-on, circ}$ of our
systems.  G95 used computer simulations of binary orbits to derive
statistical relations between the observed quantities
$\mathrm{d}_{\mathrm{proj}}$ and $v_{\mathrm{obs}}$ on the one hand and
$a$ and $v_{\rm face-on, circ}$ on the other hand.  They assumed an
eccentricity distribution of $f(\mathrm{e}) = 2\mathrm{e}$ and an isotropic
distribution of inclinations and obtained the following results:
\begin{equation}
<\mathrm{d}_{\mathrm{proj}}> = 0.91\,a\;.
\label{meana}
\end{equation}
\begin{equation}
<v_{\mathrm{obs}}> = 0.72\,v_{\rm face-on, circ}\;.
\label{meanv}
\end{equation}

However, $\mathrm{d}_{\rm proj}$ and $v_{\rm obs}$ are not uncorrelated.  For
example, if we observe a companion in the outer part of its eccentric
orbit, $\mathrm{d}_{\rm proj}$ will be larger, and $v_{\rm obs}$ smaller than
their average values.  Therefore, we cannot insert these equations
into Eq.\,\ref{msum} to obtain a statistical relation between
$\mathrm{d}_{\rm proj}$, $v_{\rm obs}$, and $M_{\rm Sys}$.

To overcome this problem, we performed more sophisticated computer
simulations.  Each simulation contains 10\,million binaries with a
fixed system mass and randomly distributed orbital parameters.  The
periods follow the distribution of periods of main-sequence stars
(Duquennoy \& Mayor \cite{D+M91}), the distribution of eccentricities
is $f(e)=2e$ and the inclinations are distributed isotropically, while
all the other parameters have uniform distributions.  The distances to
the observer are varied within a range of 143$\pm$25\,pc. We chose two
observation dates separated by a random timespan between 4 and
10~years and computed the average projected separation and orbital
velocity, in much the same way as we did for the real data.  We then
select binaries in the projected separation range from 10 to 70\,AU.
For these binaries, we compute $<v^2\cdot \mathrm{d}>/M_{\rm Sys}$.  These
simulations are repeated for different system masses.  For
$M_{\rm Sys}$ between $0.5\rm\,M_{\sun}$ and $2.5\rm\,M_{\sun}$, the
results vary from $18.4$ to $18\rm\,AU^3/(yr^2 M_{\sun})$.  This gives
us following relation:
\begin{equation}
M_{\rm Sys}\approx\frac{1}{18.2}\cdot <v^2\cdot \mathrm{d}>.
\label{msum-obs}
\end{equation}

The term on the right hand side of Eq.\,\ref{msum-obs} contains $v^2$.
Since the measured velocities and separations differ from the real
values by some unknown measurement error, this leads to an additional
bias term:
\begin{eqnarray}
<v_{\rm obs}^2\cdot \mathrm{d}_{\rm obs}>
	&=& <(v_{\rm real}+\delta v)^2\cdot (\mathrm{d}_{\rm real}+\delta
 \mathrm{d})>\nonumber\\
	&=& <v_{\rm real}^2\cdot \mathrm{d}_{\rm real}>
		+ <\delta v^2\cdot \mathrm{d}_{\rm real}>.
\label{vbias}
\end{eqnarray}

Combining Eq.\,\ref{msum-obs} and \ref{vbias}, and using $\Delta v$ as
an estimate for $\delta v$, we obtain a relation that allows us
to use the observed values to obtain an average system mass:
\begin{equation}
M_{\mathrm{Sys}} \approx
	\frac{1}{18.2}\left(<v_{\rm obs}^2\cdot \mathrm{d}_{\rm proj}>
		- <\Delta v_{\rm obs}^2\cdot \mathrm{d}_{\rm proj}>\right).
\label{meanmass}
\end{equation}

This is only an estimate for the mass in an ensemble of systems
with identical system masses.  The stars in our sample will probably
not all have the same mass.  However, we expect them to have similar
masses, and for a sufficiently large number of systems, the
uncertainties will cancel.  The result should yield a reliable average
mass for these systems.

If we use the data of all stars in our sample, Eq.\,\ref{meanmass}
yields a system mass of 2.5$\rm\,M_{\sun}$.  Excluding
\object{RX\,J1546.1-2804} lowers the result to 2.0$\rm\,M_{\sun}$.
If we further exclude \object{V 773 Tau}, \object{Elias 12}, and
\object{BD+26718B\,Aa}, we arrive at a system mass of
1.3$\rm\,M_{\sun}$.  We have reason to assume that the companion to
\object{RX\,J1546.1-2804} is a chance projected background star.
For the three other systems mentioned, the velocity is puzzling at first
sight, but is still consistent with orbital motion. Furthermore, other
possible explanations fail for \object{V 773 Tau} and
\object{Elias 12}(Sects.\,\ref{background} and \ref{unresolved}). 
Thus, it does not seem justified to exclude them from the sample, and we will
adopt 2$\rm\,M_{\sun}$ as the result for the average dynamical system mass.\\
Given the statistical uncertainties, it is difficult to estimate the error
of the system mass.  Using the standard deviation of the quantities averaged
in Eq.\,\ref{meanmass} to estimate the error of the mean yields
$\approx 0.7\rm\,M_{\sun}$.  This is in agreement with the scatter we
obtain if we exclude or include the stars mentioned in the last paragraph.\\
We conclude that the average mass of the systems in our sample is in
the range 1.3\dots2.5$\rm\,M_{\sun}$, with a most probable system mass
of 2$\rm\,M_{\sun}$.  Our result is thus consistent with the expectation
that T~Tauri stars' masses are around $M\approx 1\,M_{\sun}$ and also
with the average mass of $1.7\,M_{\sun}$ that G95 derived for their
sample.

\section{Discussion}
\label{discussion}

\subsection{Comparision with theoretical results}
We have estimated masses for the components by comparison with theoretical
PMS models for a subsample that contains 17 out of our 34 systems. In these
cases, we obtained resolved J band photometry. This spectral band is
supposed to be least affected by circumstellar excess emission and can thus
be taken as an indicator of stellar luminosity. Moreover, in these 17 
systems, there are no additional companions known, so the mass sum of the
components derived from theoretical models will match the dynamical system
masses. We placed the components of these systems
into the HRD, estimating the stellar luminosity from the resolved J band
magnitudes, assigning the optical spectral type of the system (taken
from Kenyon \& Hartmann\,\cite{Kenyon95} for Taurus-Auriga and Walter
et al.\,\cite{Walter94} for Scorpius-Centaurus) to the
primary and assuming that all components within one system are coeval.
Masses of the components were then derived using the PMS evolutionary tracks
of D'Antona \& Mazzitelli (\cite{dm98}), Swenson et al.\,(\cite{Swenson94})
and Baraffe et al.\,(\cite{Baraffe98}).\\
The mean mass obtained for this subsample from the HRD
is $0.88\,M_{\sun}$ for the D'Antona \& Mazzitelli
(\cite{dm98}) tracks and $1.17\,M_{\sun}$ for the 
Swenson et al.\,(\cite{Swenson94}) tracks. The PMS model from Baraffe et
al.\,(\cite{Baraffe98}) yields a mean mass of $1.28\,M_{\sun}$. 
The uncertainties of the mass estimates that originate from observational
data used for placing the components into the HRD are $\approx 0.1\,M_{\sun}$.
The average empirical mass derived from Eq.\,\ref{meanmass} for that
subsample\footnote{For the reasons mentioned above, the subsample does not
contain \object{RX\,J1546.1-2804}, \object{V 773 Tau}, \object{Elias\,12} and
\object{BD+26718B\,Aa}. Thus, the mean mass for the subsample is lower than
the value derived in Sect.\,\ref{systemmass}.} is $1.22\pm 0.50\,M_{\sun}$.
Within the large formal error, the predictions of all three PMS models match
our empirical result. However, our dynamical $<M>$ is much closer to the
mean masses derived using the Baraffe et al.\,(\cite{Baraffe98}) and
Swenson et al.\,(\cite{Swenson94}) tracks than to the value calculated
from the D'Antona \& Mazzitelli (\cite{dm98}) PMS model, which seems
to underestimate T~Tauri star masses. This finding has been recognized
by other authors. Bonnell et al.\,(\cite{Bonnell98}) estimated T~Tauri stars'
masses from infall velocities of accreted material and also conclude that
the empirical masses are generally larger than those predicted
by the D'Antona \& Mazzitelli (\cite{dm98}) model. A similar result
was reported by Simon et al.\,(\cite{Simon2000}), who calculated T~Tauri
stars' masses from Keplerian motion in circumstellar and circumbinary disks.

\subsection{Implications for binary statistics}
There is significant relative motion in most systems, and this motion
is in almost all cases consistent with orbital motion (Sect.\,\ref{results}).
As already pointed out by G95, this demonstrates that the large majority of
all close companions detected in the multiplicity surveys mentioned in
Sect.\,\ref{sample} really are gravitationally-bound stars. No binary
component discussed here has to be reclassified as a Herbig-Haro object
(Sect.\,\ref{herbig-haro}), and only 2 out of 34 companions may be chance
projected background stars (Sect.\,\ref{background}).\\
This finding is particularly important because there is a companion
overabundance among T~Tauri stars in the SFRs discussed here when compared
to main-sequence stars in the solar neighbourhood
(Leinert et al.\,\cite{Leinert93}, Ghez et al.\,\cite{Ghez93}, 
Ghez et al.\,\cite{Ghez97a}, K\"ohler \& Leinert\,\cite{Koehler98},
K\"ohler et al.\,\cite{Koehler2000}).
In Taurus-Auriga, almost all T~Tauri stars seem to be components of
multiple systems. To further confirm this result, K\"ohler \& Leinert
(\cite{Koehler98}) performed stellar counts in the vicinity of their
survey objects and concluded that in Taurus-Auriga, statistically
4.3 out of 44 apparent companions are projected background objects. 
K\"ohler et al.\,(\cite{Koehler2000}), in a similar way, derived a
number of 7.8 chance projections per 46 companions in Scorpius-Centaurus.
We found one candidate for a chance-projected background star
out of 21 objects in Taurus-Auriga and one candidate out of 11 objects
in Scorpius-Centaurus. The results are not directly comparable
because we have only studied the closest pairs for which chance projections
are least probable. The percentage of background stars projected by chance
among the observed companions is, however, of the same order of magnitude
in both studies. Thus, it can be concluded that chance projections
do not affect the binary statistics significantly in the SFRs discussed
here.\\
K\"ohler et al.\,(\cite{Koehler2000}) excluded six close companions
in Scorpius-Centaurus from a restricted sample. Their observed separations are
less than the strict diffraction limit $\lambda /D = 0\farcs13$ of a
3.5\,m-telescope in the K-band, so they cannot be definitely distinguished
from elongated single objects. In three of these cases, namely
\object{RX\,J1601.8-2445}, \object{RX\,J1603.9-2031B} and 
\object{RX\,J1604.3-2130B} we derive a relative velocity that is 
consistent with orbital motion (see Sect.\,\ref{results}). Thus, we propose
to classify these objects as binary systems in further studies of
multiplicity in the OB association, Scorpius-Centaurus.

\section{Summary}
\label{summary}
Based on repeated measurements of the relative astrometry in 34 close T~Tauri
binary systems, we have reproduced the results given by G95 and extended their
work to a larger number of binaries and particularly to a much longer
timespan of up to 10 years. We showed that in most systems significant
relative motion of the components has occured. In almost all cases this
relative motion can be explained by orbital motion. In only two systems the
observed motion may be the result of the proper motion of a T~Tauri star that
is accidentally projected in the close vicinity of a background star.\\
From the short pieces of orbit available at the moment (up to $20^{\circ}$ in
position angle), we derive a mean dynamical system mass of
$2.0\pm 0.7\,M_{\odot}$ for our sample.
This mass is consistent with the predictions of current sets of
PMS evolutionary models within the uncertainties. The large formal error of
this mean mass does not allow a significant discrimination between different
models, but we draw the tentative conclusion that the masses predicted by the
D'Antona \& Mazzitelli\,(\cite{dm98}) model may be systematically too low.\\
The result that orbital motion can be detected in most systems discussed
here indicates that the ``companions'' found in previous multiplicity
surveys really are gravitationally bound stars. This is a further
confirmation of the binary overabundance in Taurus-Auriga and
Scorpius-Centaurus compared to nearby main sequence stars. Furthermore,
the detection of orbital motion allows a definite classification
of three objects with very close separations as stellar companions.

\begin{acknowledgements}
We thank the staff at ESO La Silla and Calar Alto for their support
during several observing runs. In particular we are grateful to Andreas
Eckart and Klaus Bickert for their support in observing with the SHARP I
camera. We modified a program written by Sabine Frink to carry out the
computer simulations described in Sect.\,\ref{systemmass}.
The authors appreciate fruitful discussions with Michal Simon, and thank
the anonymous referee for fair and constructive criticism.
\end{acknowledgements}

\appendix

\section{Overview of observations}
Table\,\ref{obs-table} presents an overview of all measurements
of the relative astrometry in T~Tauri binary systems that form
the database for the present study. In Fig.\,\ref{orbitplots} these
measurements are plotted as a function of epoch for the individual
systems in cartesian and polar coordinates.

\begin{table*}
\caption{\label{obs-table} Relative astrometry of the components in
 young binary systems measured at different epochs. If no reference
 is given the results were obtained by the authors.}
\begin{tabular}{lllllll}
\hline
System & Date & Filter & Flux ratio $I_2/I_1$ &
 PA/$^{\circ}$ & $d_{\mathrm{proj}}$/mas & Reference \\ 
\hline
V\,773\,Tau & 3 Oct 1990 & K & & 295 $\pm$ 4 & 112 $\pm$ 1 & G95 \\
            & 21 Sep 1991 & K & 0.13 $\pm$ 0.04 & 295 $\pm$ 3 &
              170 $\pm$ 10 & \\
            & 10 Oct 1992 & K & & 307 $\pm$ 5 & 120 $\pm$ 20 & G95 \\
            & 5 Oct 1993 & J & 0.113 $\pm$ 0.005 &
              308.9 $\pm$ 1.4 & 111 $\pm$ 4 & \\
            & 25 Nov 1993 & K & & 304 $\pm$ 8 & 92 $\pm$ 8 & G95 \\
            & 19 Oct 1994 & K & & 318 $\pm$ 2 & 65 $\pm$ 1 & G95 \\
            & 29 Oct 1994 & UBVRI (HST) & & 321.9 $\pm$ 2.7 & 63 $\pm$ 2 &
               Ghez et al.\,(\cite{Ghez97b}) \\
LkCa\,3     & 6 Dec 1990 & K & 0.5 & 78 $\pm$ 1 & 470 $\pm$ 40 & \\
            & 5 Oct 1993 & J & 0.62 $\pm$ 0.02 & 74.5 $\pm$ 0.7 &
            494 $\pm$ 13 & \\
            & 29 Sep 1996 & H & 0.885 $\pm$ 0.011 & 74.2 $\pm$ 0.3 &
            485 $\pm$ 4 & \\
            & 19 Nov 1997 & K & 0.755 $\pm$ 0.007 &  73.4 $\pm$ 0.1 &
 485 $\pm$ 4 & \\
FO\,Tau     & 19 Sep 1991 & K & 0.92 $\pm$ 0.04 & 180 $\pm$ 4 &
             165 $\pm$ 5 & \\
            & 18 Oct 1991 & K & & 181.7 $\pm$ 0.6 & 161 $\pm$ 1 & G95 \\
            & 5 Oct 1993 & J & 0.30 $\pm$ 0.01 & 182.0 $\pm$ 0.9 &
              183 $\pm$ 5 & \\
            & 20 Oct 1994 & K & & 190.6 $\pm$ 0.4 & 153 $\pm$ 2 & G95 \\
            & 14 Dec 1994 & K & 0.72 $\pm$ 0.02 & 189.7 $\pm$ 0.9 &
            159 $\pm$ 4 & \\
            & 19 Dec 1994 & K & & 191.2 $\pm$ 0.4 & 154 $\pm$ 2 & G95 \\
            & 9 Oct 1995 & K & 0.648 $\pm$ 0.018 & 193.1 $\pm$ 1.3 & 
            156 $\pm$ 4 & \\
            & 27 Sep 1996 & K & 0.628 $\pm$ 0.019 & 194.3 $\pm$ 0.9 &
            152 $\pm$ 4 & \\
            & 29 Nov 1996 & J & 0.55 $\pm$ 0.05 & 200.0 $\pm$ 1.1 &
            143 $\pm$ 4 & \\
            & 16 Nov 1997 & H & 0.698 $\pm$ 0.014 & 198.9 $\pm$ 0.8 &
            150 $\pm$ 5 & \\
CZ\,Tau     & 19 Mar 1992 & K & 0.46 $\pm$ 0.03 & 84 $\pm$ 3 &
            330 $\pm$ 10 & \\
            & 28 Sep 1996 & K & 0.183 $\pm$ 0.004 & 90.3 $\pm$ 0.3 &
            306 $\pm$ 4 & \\
            & 29 Nov 1996 & J & 0.120 $\pm$ 0.005 & 89.7 $\pm$ 1.1 &
            320 $\pm$6 & \\
FS\,Tau     & 20 Sep 1989 & K & & 60 $\pm$ 5 & 265 $\pm$ 5 & 
              Simon et al.\,(\cite{Simon92}) \\
            & 25 Jan 1996 & RI (HST) & & 84.0 $\pm$ 1.5 & 239 $\pm$ 5 & 
              Krist et al.\,(\cite{Krist98}) \\
            & 28 Sep 1996 & H & 0.183 $\pm$ 0.008 & 84.4 $\pm$ 1.6 &
            238 $\pm$ 4 & \\
            & 29 Nov 1996 & J & 0.188 $\pm$ 0.007 & 83.3 $\pm$ 3.0 &
             265 $\pm$ 15 & \\
            & 17 Nov 1997 & K & 0.138 $\pm$ 0.005 & 83.3 $\pm$ 1.5 &
            248 $\pm$ 5 & \\
FW\,Tau     & 17 Oct 1989 & K & 1.00 $\pm$ 0.01 & 158 $\pm$ 2 &
            151$\pm$ 1 & \\
            & 31 Mar 1990 & K & & 160 $\pm$ 5  & 160$\pm$ 20 &
              Simon et al.\,(\cite{Simon92}) \\
            & 13 Dec 1994 & K & 0.61 $\pm$ 0.10 & 187.4 $\pm$ 1.6 &
            98 $\pm$ 6 & \\
            & 9 Oct 1995 & K & 1.00 $\pm$ 0.05 &  188.5 $\pm$ 3.7 &
            78 $\pm$ 5 & \\
            & 27 Sep 1996 & H & 0.76 $\pm$ 0.10 & 190.1 $\pm$ 3.3 &
            71 $\pm$ 4 & \\
LkH$\alpha$ 331 & 29 Oct 1991  & K & 0.73 $\pm$ 0.04 & 290 $\pm$ 4  &
                300 $\pm$ 10 & \\
                & 26 Jan 1994  & J & 0.706 $\pm$ 0.023 & 286.2 $\pm$ 0.9 &
                279 $\pm$ 4 & \\
                & 9 Oct 1995   & K & 0.66 $\pm$ 0.03 & 286.6 $\pm$ 0.8 &
                279 $\pm$ 4 & \\
                & 29 Sep 1996  & H & 0.70 $\pm$ 0.1 & 287.8 $\pm$ 0.2 &
                278 $\pm$ 4 & \\
XZ\,Tau         & 17 Oct 1989  & K & 0.35 $\pm$ 0.03 & 154 $\pm$ 3 &
                300 $\pm$ 20 & \\
                & 10 Oct 1991  & K & & 151 $\pm$ 2 & 310 $\pm$ 10 & G95 \\
                & 27 Jan 1994  & J & 1.51 $\pm$ 0.03 & 146.1 $\pm$ 0.6 &
                307 $\pm$ 4 & \\
                & 28 Jan 1994  & K & 0.41 $\pm$ 0.01 & 146.8 $\pm$ 1.0 &
                306 $\pm$ 5 & \\
                & 19 Dec 1994  & K & & 147.0 $\pm$ 0.4 & 296 $\pm$ 2 & G95 \\
                & 5 Jan 1995   & RI (HST) & & 147.8 $\pm$ 0.25 &
                307 $\pm$ 9 & Krist et al.\,(\cite{Krist97}) \\
                & 29 Sep 1996  & K & & 145.4 $\pm$ 0.3  &
                299 $\pm$ 4 & \\
                & 30 Nov 1996  & J & 3.54 $\pm$ 0.31 & 145.7 $\pm$ 0.7  &
                309 $\pm$ 10 & \\
                & 19 Nov 1997  & K & 0.316 $\pm$ 0.007 & 140.9 $\pm$ 0.5 &
                302 $\pm$ 4 \\
HK\,Tau\,G2     & 27 Sep 1991  & K & 0.88 $\pm$ 0.03 & 300 $\pm$ 4 &
                190 $\pm$ 10 & \\
                & 19 Oct 1991  & K & & 304.1 $\pm$ 0.5 & 163 $\pm$ 5 & G95 \\
                & 26 Jan 1994  & J & 0.764 $\pm$ 0.023 & 302.5 $\pm$ 0.5 &
                176 $\pm$ 4 & \\
                & 27 Jan 1994  & H & 0.85 $\pm$ 0.12 & 301.5 $\pm$ 1.7 &
                180 $\pm$ 10 & \\
                & 8 Dec 1994   & V & & 302.4 $\pm$ 0.6 & 178 $\pm$ 2 &
                  Simon et al.\,(\cite{Simon96}) \\
                & 19 Dec 1994 & K & & 301.2 $\pm$ 0.6 & 181 $\pm$ 2 & G95 \\
                & 9 Oct 1995  & K & 0.85 $\pm$ 0.06 & 300.3 $\pm$ 1.0 &
                188 $\pm$ 4 & \\
                & 27 Sep 1996 & H & 0.76 $\pm$ 0.05 & 299.6 $\pm$ 0.6 &
                193 $\pm$ 4 & \\
                & 19 Nov 1997 & K & 0.587 $\pm$ 0.017 & 298.3 $\pm$ 0.6 &
                214 $\pm$ 4 & \\ 
\hline
\end{tabular}
\end{table*}
\clearpage

\setcounter{table}{0}
\begin{table*}
\caption{continued}
\begin{tabular}{lllllll}
\hline
System & Date & Filter & Flux ratio $I_2/I_1$ & PA/$^{\circ}$ &
 $d_{\mathrm{proj}}$/mas & Reference \\ 
\hline
GG Tau     & 2 Nov 1990   & K & 0.64 $\pm$ 0.1 & 9 $\pm$ 2 & 255 $\pm$ 10 & \\
           & 21 Oct 1991  & K & & 2 $\pm$ 1 & 260 $\pm$ 10 & G95 \\
           & 23 Dec 1994  & IJHK & & 3 $\pm$ 2 & 260 $\pm$ 10 &
             Roddier et al.\,(\cite{Roddier96}) \\
           & 27 Jan 1994  & J & 0.543 $\pm$ 0.004 & 357.8 $\pm$ 0.4 &
           246 $\pm$ 4 & \\
           & 25 Jul 1994  & UBVRI (HST) & & 358.8 $\pm$ 0.5 & 250 $\pm$ 2 & 
              Ghez et al.\,(\cite{Ghez97b}) \\
           & 24 Sep 1994  & K & & 357   $\pm$ 2   & 258 $\pm$ 4 & G95 \\
           & 18 Oct 1994  & K & & 0.9   $\pm$ 0.5 & 242 $\pm$ 3 & G95 \\
           & 8 Oct 1995   & K & 0.54 $\pm$ 0.02 & 356.9 $\pm$ 0.7 &
           247 $\pm$ 4 & \\
           & 29 Sep 1996  & H & 0.556 $\pm$ 0.007 & 355.5 $\pm$ 0.4 &
           245 $\pm$ 4 & \\
           & 16 Nov 1997  & K & 0.564 $\pm$ 0.004 & 353.6 $\pm$ 0.1 &
           247 $\pm$ 5 & \\
           & 10 Oct 1998  & K & 0.476 $\pm$ 0.005 & 350.7 $\pm$ 0.4 &
           260 $\pm$ 4 & \\
UZ\,Tau/w  & 8 Oct 1990   & K & & 0 $\pm$ 8 & 340 $\pm$ 60  & 
              Simon et al.\,(\cite{Simon92}) \\
           & 10 Nov 1990  & K & & 357.4 $\pm$  0.3 & 359 $\pm$ 2   & G95 \\
           & 26 Jan 1994  & J & 0.76 $\pm$ 0.07 & 1.3 $\pm$ 1.0    &
           358 $\pm$ 10  & \\
           & 24 Jul 1994  & UBVRI (HST) & & 0.1 $\pm$ 0.1    &
           368 $\pm$ 1   & Ghez et al.\,(\cite{Ghez97b}) \\
           & 19 Dec 1994  & K & & 1.3 $\pm$ 0.4    & 360 $\pm$ 3   & G95 \\
           & 29 Sep 1996  & K & 0.66 $\pm$ 0.02 & 1.3 $\pm$ 0.2    &
           366 $\pm$ 4   & \\
GH\,Tau    & 19 Oct 1991  & K & & 118.4 $\pm$ 0.1 & 314 $\pm$ 1 & G95 \\
           & 27 Oct 1991  & K & 0.91 $\pm$ 0.04 & 120 $\pm$ 1 &
           350 $\pm$ 10 & \\
           & 27 Jan 1994  & J & 0.89 $\pm$ 0.01 & 116.4 $\pm$ 0.3 &
           313 $\pm$ 3 & \\
           & 19 Dec 1994  & K & & 116.0 $\pm$ 0.4 & 307 $\pm$ 2 & G95 \\
           & 29 Sep 1996  & H & 1.03 $\pm$ 0.10 & 111.4 $\pm$ 1.8 &
           300 $\pm$ 10 & \\
Elias\,12  & 10 Nov 1990 & K & & 332.5 $\pm$ 0.2 & 371 $\pm$ 1 & G95 \\
           & 27 Oct 1991 & K & 0.45 $\pm$ 0.02 & 333.9 $\pm$ 0.8 &
           411 $\pm$ 10 & \\
           & 5 Jan 1993  & H & 0.54 $\pm$ 0.01 & 332.4 $\pm$ 0.8 &
           340 $\pm$ 10 & \\
           & 26 Nov 1993 & K & & 334 $\pm$ 6 & 340 $\pm$ 20 & G95 \\
           & 27 Jan 1994 & J & 0.579 $\pm$ 0.022 & 328.5 $\pm$ 0.9 &
           347 $\pm$ 5 & \\
           & 27 Aug 1994 & V & & 328.5 $\pm$ 0.4 & 335.5 $\pm$ 2  &
              Simon et al.\,(\cite{Simon96}) \\
           & 24 Sep 1994 & K & & 326.9 $\pm$ 0.9 & 336 $\pm$ 3 & G95 \\
           & 18 Oct 1994 & K & & 328.1 $\pm$ 0.6 & 329 $\pm$ 4 & G95 \\
           & 4 Feb 1995 & V & & 328.8 $\pm$ 0.4 & 337.5 $\pm$ 2 & 
              Simon et al.\,(\cite{Simon96}) \\
           & 29 Sep 1996 & H & 0.58 $\pm$ 0.04 & 325.2 $\pm$ 0.4 &
           318 $\pm$ 4 & \\
IS\,Tau    & 19 Oct 1991 & K & & 92   $\pm$ 2 & 221 $\pm$ 4 & 
             Ghez et al.\,(\cite{Ghez93}) \\
           & 10 Jan 1993 & K  & 0.21 $\pm$ 0.02 & 86   $\pm$ 8 &
            210 $\pm$ 20 & \\
           & 26 Jan 1994 & J & 0.32 $\pm$ 0.01 & 93.5 $\pm$ 0.5 &
            238 $\pm$ 4 & \\
           & 9 Oct 1995  & K & 0.165 $\pm$ 0.004 & 96.3 $\pm$ 1.0 &
             235 $\pm$ 4 & \\
           & 28 Sep 1996 & H & 0.20 $\pm$ 0.01 & 97.0 $\pm$ 0.2 &
             223 $\pm$ 4 & \\
           & 29 Nov 1996 & J & 0.21 $\pm$ 0.01 & 95.3 $\pm$ 1.5 &
             210 $\pm$ 9 & \\
IW\,Tau    & 28 Oct 1991 & K & 0.91 $\pm$ 0.04 & 177 $\pm$ 2 &
           270 $\pm$ 20 & \\
           & 27 Jan 1994 & J & 1.24 $\pm$ 0.04 & 176.0 $\pm$ 1.0 &
           278 $\pm$ 5 & \\
           & 28 Sep 1996 & H & 1.30 $\pm$ 0.15 & 179.4 $\pm$ 0.7 &
           288 $\pm$ 7 & \\
LkH$\alpha$\,332/G2 & 27 Oct 1991 & K & 0.60 $\pm$ 0.05 & 243 $\pm$ 2 &
                    300 $\pm$ 10 & \\
                    & 27 Jan 1994 & J & 0.458 $\pm$ 0.003 & 236.4 $\pm$ 0.3 &
                    260 $\pm$ 4 & \\
                    & 15 Dec 1994 & K & 0.530 $\pm$ 0.017 & 237.5 $\pm$ 1.0 &
                    252 $\pm$ 4 & \\
                    & 28 Sep 1996 & H & 0.411 $\pm$ 0.012 & 240.1 $\pm$ 0.7 &
                    239 $\pm$ 4 & \\
                    & 17 Nov 1997 & K & 0.465 $\pm$ 0.022 & 241.6 $\pm$ 0.4 &
                    234 $\pm$ 4 & \\
LkH$\alpha$\,332/G1 & 4 Oct 1990 & K & & 77.6 $\pm$ 0.4 & 208 $\pm$ 2 & G95 \\
                    & 27 Oct 1991 & K & 0.58 $\pm$ 0.03 & 85 $\pm$ 2 &
                    230 $\pm$ 20 & \\
                    & 25 Nov 1993 & K & & 83 $\pm$ 2 & 230 $\pm$ 10 & G95 \\
                  & 18 Oct 1994 & K & & 82.7 $\pm$ 0.4 & 224 $\pm$ 2 & G95 \\
                    & 12 Dec 1994 & K & 0.555 $\pm$ 0.006 & 83.8 $\pm$ 0.3 &
                    228 $\pm$ 4 & \\
                    & 28 Sep 1996 & H & 0.560 $\pm$ 0.009 & 86.3 $\pm$ 0.4 &
                    228 $\pm$ 4 & \\
                    & 29 Nov 1996 & J & 0.511 $\pm$ 0.026 & 87.4 $\pm$ 0.3 &
                    236 $\pm$ 5 & \\
                    & 22 Nov 1997 & K & 0.51 $\pm$ 0.02 & 91.0 $\pm$ 0.3 &
                    230 $\pm$ 4 & \\
LkH$\alpha$\,332    &  27 Oct 1991 & K & 0.23 $\pm$ 0.03 & 204 $\pm$ 2 &
                    360 $\pm$ 10 & \\ 
                    &  27 Jan 1994 & J & 0.80 $\pm$ 0.04 & 204.5 $\pm$ 1.0 &
                    313 $\pm$ 11 & \\
                    &  28 Sep 1996 & H & 0.485 $\pm$ 0.015 & 204.8 $\pm$ 1.3
                       & 325 $\pm$ 10 & \\
                    &  22 Nov 1997 & K & 0.46 $\pm$ 0.01 & 203.5 $\pm$ 0.3 &
                    343 $\pm$ 4 & \\
\hline
\end{tabular}
\end{table*}
\clearpage

\setcounter{table}{0}
\begin{table*}
\caption{continued}
\begin{tabular}{lllllll}
\hline
System & Date & Filter & Flux ratio $I_2/I_1$ & PA/$^{\circ}$ &
  $d_{\mathrm{proj}}$/mas & Reference \\ 
\hline
BD+26 718B - Aa   & 16 Sep 1994  & K & & 320.1 $\pm$ 0.5 & 496 $\pm$ 3  & 
  K\"ohler \& Leinert (\cite{Koehler98}) \\
                  & 27 Sep 1996  & K & & 320.0 $\pm$ 0.1 & 474 $\pm$ 3  &
  K\"ohler \& Leinert (\cite{Koehler98}) \\
                  & 18 Nov 1997  & K & 0.178 $\pm$ 0.004 & 319.9 $\pm$ 0.4
                  & 454 $\pm$ 5  & \\
BD+26 718B - B    & 16 Sep 1994  & K & & 136.8 $\pm$ 9.0 & 166 $\pm$ 7 &
  K\"ohler \& Leinert (\cite{Koehler98}) \\
                  & 27 Sep 1996  & K & & 133.4 $\pm$ 0.2 & 155 $\pm$ 3 &
  K\"ohler \& Leinert (\cite{Koehler98}) \\
                  & 18 Nov 1997  & K & 0.310 $\pm$ 0.006 & 135.2 $\pm$ 0.2
                  & 171 $\pm$ 4 & \\
BD+17 724  & 8 Oct 1995 & K & & 208.3 $\pm$ 9.7 & 98 $\pm$ 11 &
  K\"ohler \& Leinert (\cite{Koehler98}) \\
           & 27 Sep 1996 & K & & 193.9 $\pm$ 2.1 & 84 $\pm$ 3 &
  K\"ohler \& Leinert (\cite{Koehler98}) \\
           & 18 Nov 1997  & H &   & 190.7 $\pm$ 1.8 & 87 $\pm$ 5 & \\
HM\,Anon   & 4 May 1994  & K &  & 236 $\pm$ 5 & 270 $\pm$ 30  & 
       Ghez et al.\,(\cite{Ghez97a}) \\
           & 5 May 1998  & H & 0.240 $\pm$ 0.006 & 243.6 $\pm$ 1.0 &
            256 $\pm$ 7  & \\
           &  & J & 0.205 $\pm$ 0.004 \\
HN\,Lup    & 27 Apr 1994 & K & & 359 $\pm$ 2  & 240 $\pm$ 10 &
       Ghez et al.\,(\cite{Ghez97a}) \\
           & 7 May 1998 & H & 0.628 $\pm$ 0.013 & 351.3 $\pm$ 0.2 &
            247 $\pm$ 4 & \\
           &            & J & 0.831 $\pm$ 0.044 & \\
RX\,J1546.1-2804 & 1 May 1994 & K & & 41.2 $\pm$ 0.5 & 110 $\pm$ 3 &
   K\"ohler et al.\,(\cite{Koehler2000}) \\
                 & 7 May 1998 & H & 0.537 $\pm$ 0.016 & 318.7 $\pm$ 0.5 &
                 88 $\pm$ 4 & \\
RX\,J1549.3-2600 & 1 May 1994 & K & & 320.3 $\pm$ 0.4 & 164 $\pm$ 4 &
    K\"ohler et al.\,(\cite{Koehler2000}) \\
                 & 7 May 1998  & H & 0.443 $\pm$ 0.003 & 322.2 $\pm$ 0.2 &
                 163 $\pm$ 4 & \\
RX\,J1600.5-2027 & 2 May 1994 & K & & 171.7 $\pm$ 0.5 & 189 $\pm$ 4 &
   K\"ohler et al.\,(\cite{Koehler2000}) \\
                 & 7 May 1998 & H & 0.735 $\pm$ 0.021 & 165.3 $\pm$ 0.3 &
                 200 $\pm$ 4 & \\
RX\,J1601.7-2049 & 2 May 1994  & K & & 324.7 $\pm$ 0.9 & 205 $\pm$ 4 &
   K\"ohler et al.\,(\cite{Koehler2000}) \\
                 & 7 May 1998 & H & 0.504 $\pm$ 0.012 & 320.5 $\pm$ 0.2 & 
                 205 $\pm$ 4 & \\
RX\,J1601.8-2445 & 10 Jul 1995  & K & & 289.6 $\pm$ 10  & 76 $\pm$ 5  &
    K\"ohler et al.\,(\cite{Koehler2000}) \\
                 & 7 May 1998 & H & 0.595 $\pm$ 0.030 & 280.4 $\pm$ 1.5 &
                 108 $\pm$ 4 & \\
RX\,J1603.9-2031B  &   2 May 1994 & K & & 140.9 $\pm$ 0.6 & 121 $\pm$ 4 &
    K\"ohler et al.\,(\cite{Koehler2000}) \\
                 & 7 May 1998 & H & 0.666 $\pm$ 0.022 & 113.3 $\pm$ 1.3 &
                 96 $\pm$ 4 & \\
RX\,J1604.3-2130B  & 1 May 1994  & K & & 327.4 $\pm$ 0.8 & 82 $\pm$ 4 &
    K\"ohler et al.\,(\cite{Koehler2000}) \\
                   &  7 May 1998 & H & 0.471 $\pm$ 0.004 & 342.3 $\pm$ 0.7 &
                   88 $\pm$ 4 & \\
NTTS\,155808-2219  & 2 May 1994 & K & & 313.7 $\pm$ 1.2 & 193 $\pm$ 5 &
   K\"ohler et al.\,(\cite{Koehler2000}) \\
                   & 7 May 1998 & H & 0.575 $\pm$ 0.008 & 317.5 $\pm$ 0.8 &
                    213 $\pm$ 4 & \\
                   &            & J & 0.585 $\pm$ 0.019 & \\
NTTS 160735-1857   & 2 May 1994 & K & & 84.1 $\pm$ 0.3 & 299 $\pm$ 4  &
     K\"ohler et al.\,(\cite{Koehler2000}) \\
                   & 8 May 1994 & H & 0.584 $\pm$ 0.021 & 76.9 $\pm$ 1.0 &
                   300 $\pm$ 7 &  \\
                   &            & J & 0.656 $\pm$ 0.019 \\
NTTS\,155913-2233  & 8 Jul 1990 & K & & 347.7 $\pm$ 0.7 & 283 $\pm$ 4 & G95 \\
                   & 4 Jul 1993 & K & & 345 $\pm$ 2 & 292 $\pm$ 8 & G95 \\
                   & 1 May 1994 & K & & 345.6 $\pm$ 0.4 &  301 $\pm$ 4 &
   K\"ohler et al.\,(\cite{Koehler2000}) \\  
                   & 7 May 1998 & H & 0.577 $\pm$ 0.016 & 337.6 $\pm$ 0.4 &
                   318 $\pm$ 4 & \\
                   &            & J & 0.548 $\pm$ 0.014 \\
NTTS 160946-1851   & 5 Aug 1990  & K & & 160.4 $\pm$ 0.6 & 208 $\pm$ 2  & G95 \\
                   & 4 May 1991  & K & & 162 $\pm$ 3 & 212 $\pm$ 5 & G95 \\
                   & 12 May 1992 & K & & 160 $\pm$ 4 & 210 $\pm$ 20  & G95 \\
                   & 25 Jul 1992 & K & & 158.6 $\pm$ 0.5 & 212 $\pm$ 5 & G95 \\
                   & 1 Jul 1993 & K & & 159 $\pm$ 1 & 209 $\pm$ 2 & G95 \\
                   & 2 May 1994 & K & & 161.9 $\pm$ 0.4 & 203 $\pm$ 6 &
   K\"ohler et al.\,(\cite{Koehler2000}) \\  
                   & 8 May 1998 & H & 0.194 $\pm$ 0.009 & 161.4 $\pm$ 1.5 &
                    220 $\pm$ 13 & \\
                   &            & J & 0.210 $\pm$ 0.007 & \\
\hline
\end{tabular}
\end{table*}
\clearpage

\begin{figure*}
\resizebox{\hsize}{!}{\includegraphics{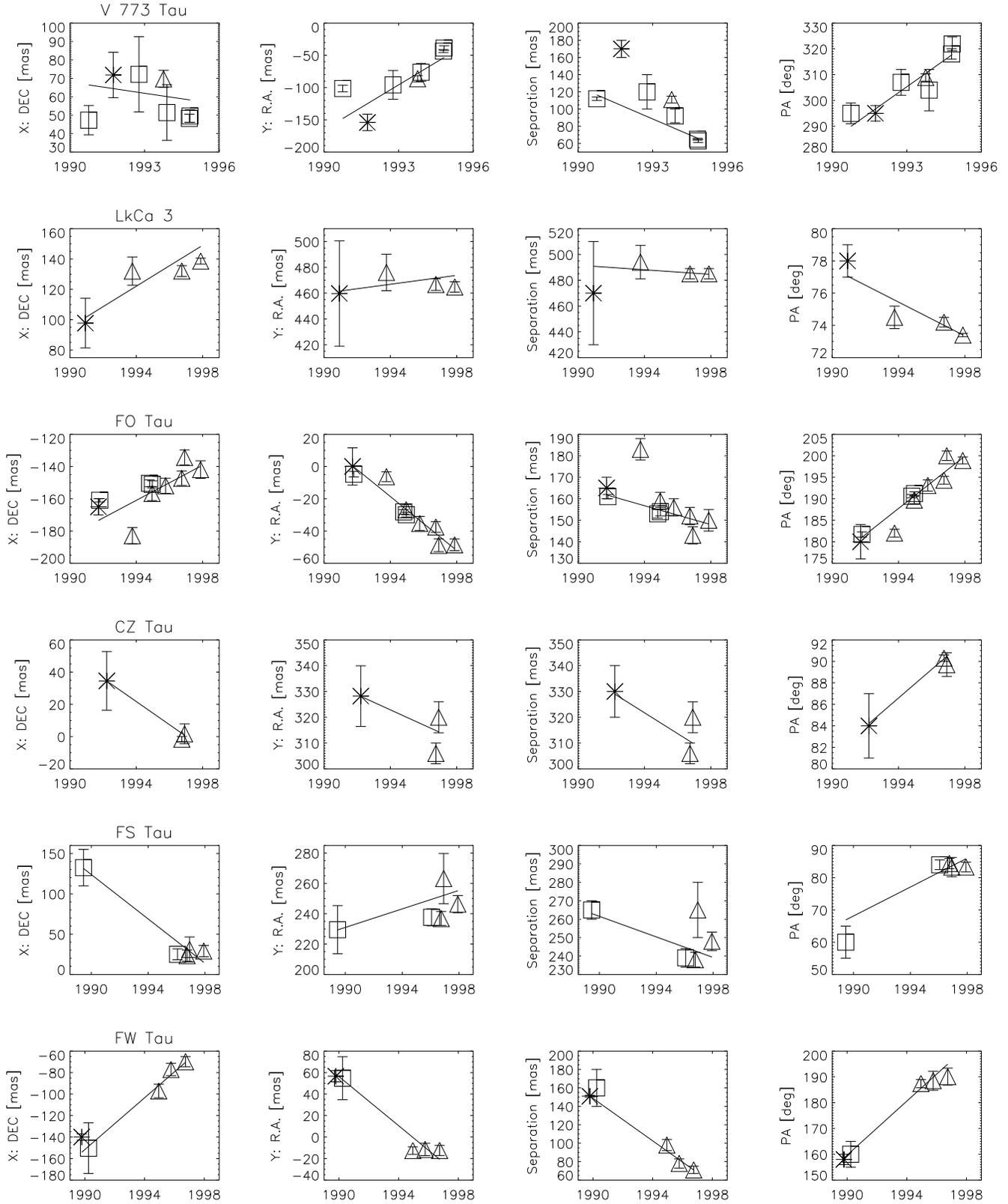}}
  \caption{\label{orbitplots} Relative astrometry of the components in
  T~Tauri binary systems in cartesian coordinates (first and second column)
  and polar coordinates (third and fourth column). The solid lines
  indicate the results of weighted linear fits to this data. Triangles
  denote our new data presented in this paper. Asterisks refer to
  measurements of ``first epoch'' that coincide with the detection of
  the companions. They have already been published (Leinert et
  al.\,\cite{Leinert93} and K\"ohler \& Leinert\,\cite{Koehler98}). Squares
  indicate data points taken from literature, in most cases from G95
 (see Table\,\ref{obs-table} for reference).}
\end{figure*}
\clearpage

\setcounter{figure}{0}
\begin{figure*}
 \resizebox{\hsize}{!}{\includegraphics{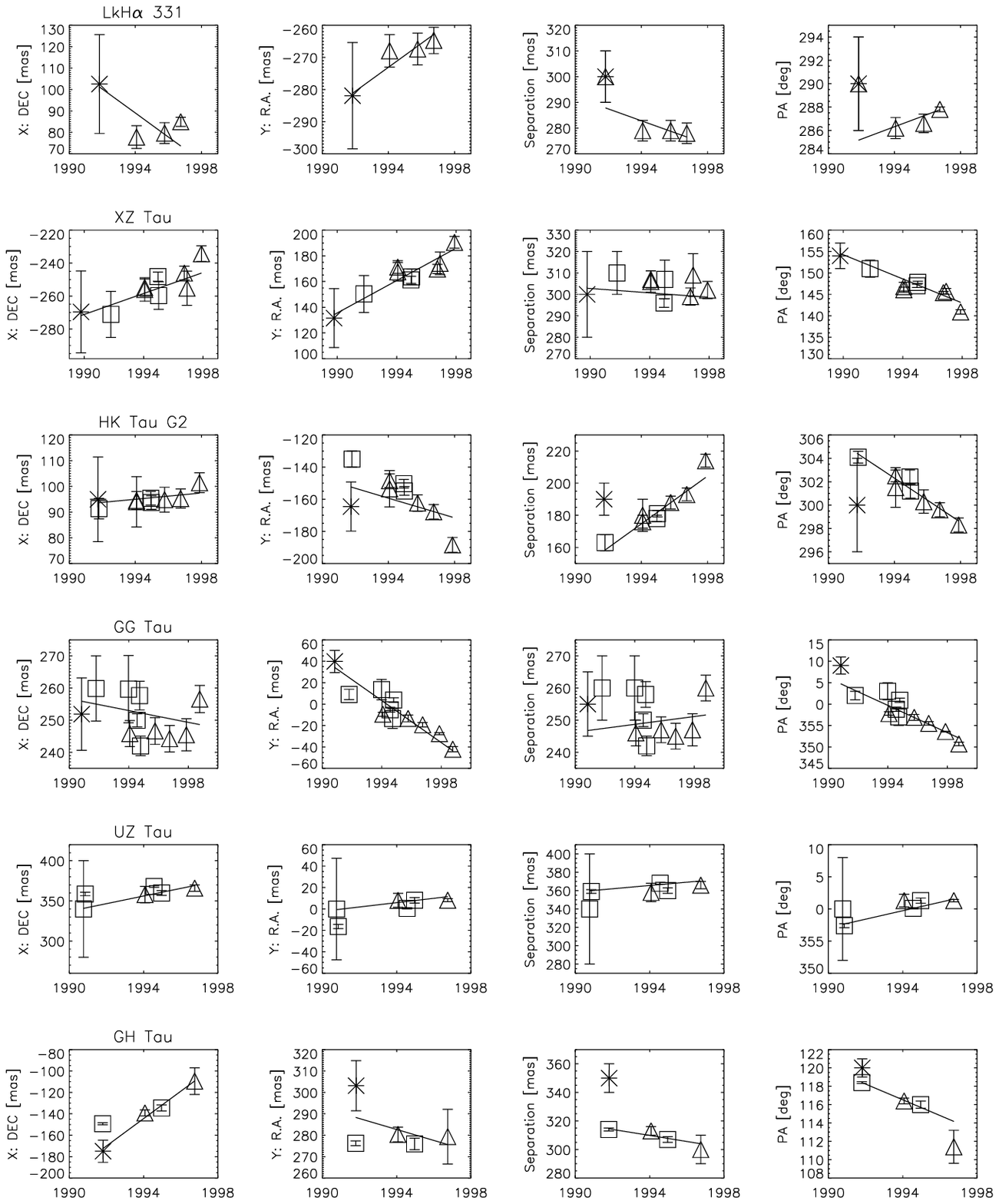}}
 \caption{continued}
\end{figure*}
\clearpage

\setcounter{figure}{0}
\begin{figure*}
 \resizebox{\hsize}{!}{\includegraphics{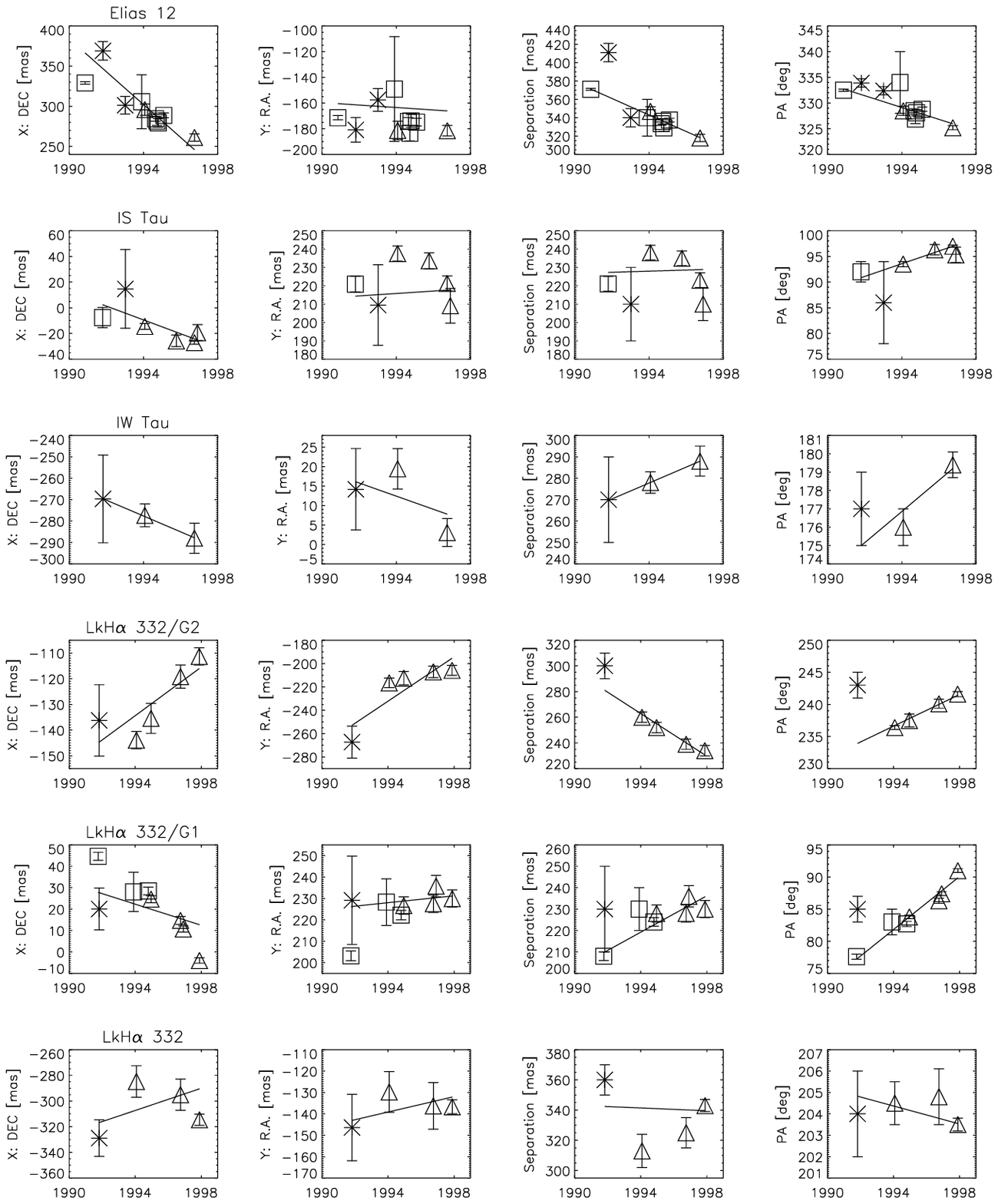}}
 \caption{continued}
\end{figure*}
\clearpage

\setcounter{figure}{0}
\begin{figure*}
 \resizebox{\hsize}{!}{\includegraphics{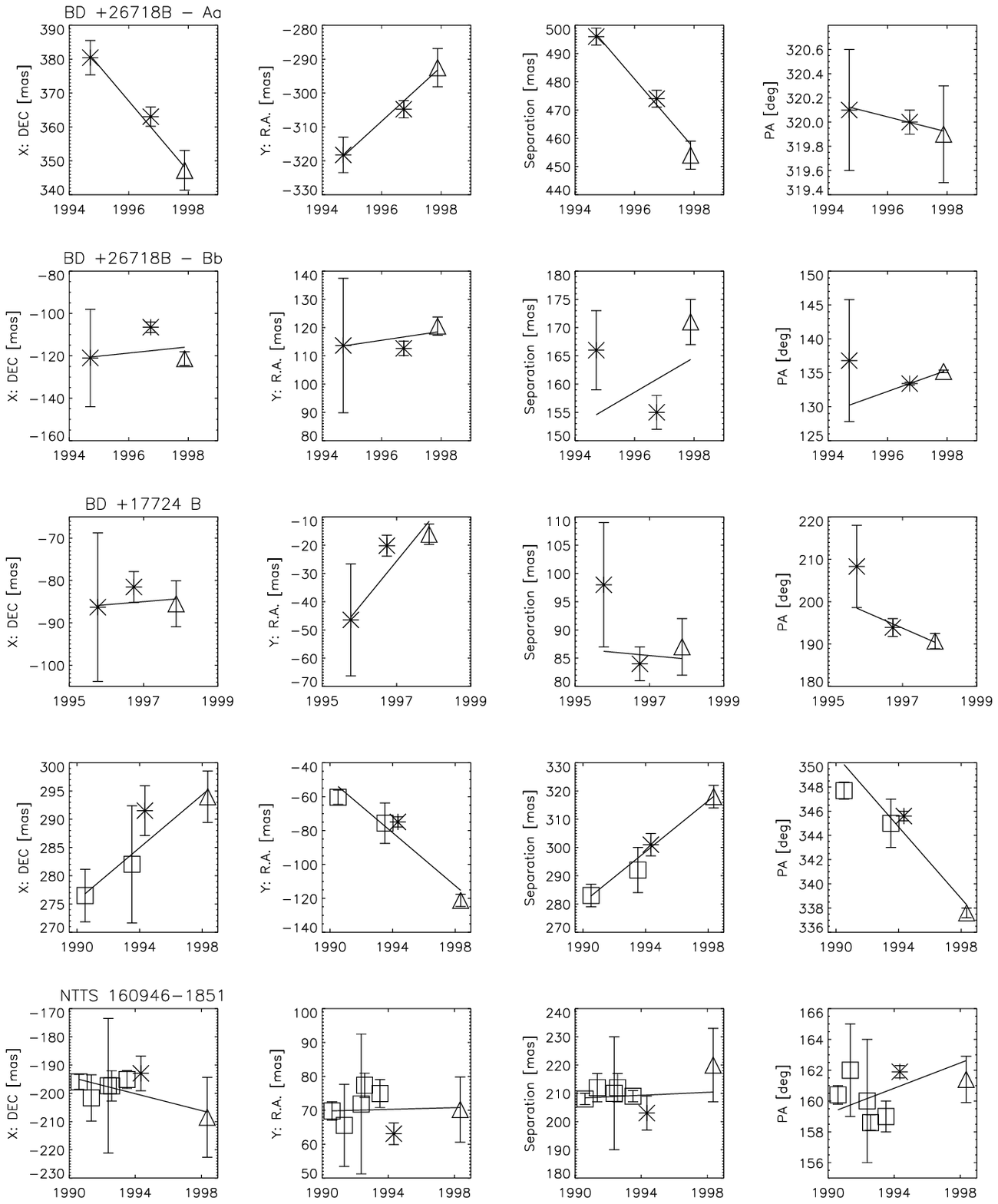}}
 \caption{continued}
\end{figure*}
\clearpage

\end{document}